\documentclass[11pt]{article}

\usepackage{lmodern} % use Latin Modern for that LaTeX look

\usepackage[round]{natbib}
\usepackage{graphicx, url}
\usepackage{amsmath}
\usepackage{amsfonts}
\usepackage{amssymb}
\usepackage{pdfsync}

\usepackage{authblk}
\usepackage{anysize}
\marginsize{2cm}{2.5cm}{2.5cm}{2cm}
\DeclareMathOperator{\vect}{vec}
\usepackage{algorithm2e}

\newtheorem{Remark}{Remark}[section]

\newtheorem{Definition}{Definition}[section]

\newcommand{\mathbold}[1]{\mbox{\boldmath $\bf#1$}}

\usepackage[textwidth=3cm, textsize=footnotesize]{todonotes} %%%%new command for comments

\author[]{Konstantinos Fokianos}

\affil[]{Department of Mathematics \& Statistics, University of Cyprus}

{
	\makeatletter
	\renewcommand\AB@affilsepx{: \protect\Affilfont}
	\makeatother
	
	\affil[ ]{Email}
	
	\makeatletter
	\renewcommand\AB@affilsepx{, \protect\Affilfont}
	\makeatother
	
	\affil[]{fokianos@ucy.ac.cy}
}

\title{Multivariate Count Time Series Modelling}
\date{\today}

\begin{document}
\maketitle

\begin{abstract}
\noindent
We review autoregressive models  for the analysis of  multivariate count time series. In doing so, we discuss the choice 
of a suitable distribution for  a  vectors of count random variables. This review focus on three main approaches taken for multivariate count time series analysis: (a) integer autoregressive processes, (b) parameter-driven models  and (c) observation-driven models. The aim of this 
work is to highlight some recent methodological developments  and  propose some potentially useful research topics.
\end{abstract}
\noindent
{\textbf{Keywords:} {\small auto-correlation, covariates, copula, estimation, multivariate count distribution, prediction}

\newpage 

\section{Introduction}
\label{sec:Introduction}

This work  reviews three  main approaches that have been put forward  for analysis and inference of 
multivariate count time series.
By now there is an extensive literature for modeling univariate count time series, see the recent volume by \cite{Davisetal(2016)}
and the review article by \citet{Davisetal(2021)}, for example.  Theoretical and methodological development for   multivariate count time series 
is still on-going research area; 
see  \citet{Paul2008} for a medical application,
\citet{PedeliandKarlis(2013)} for a financial study  and more recently \citet{Ravishankeretal(2015)}, for a marketing application,
and \citet{Livseyetal(2018)} for an environmental study.
The interested reader is referred to the review paper by  \citet{Karlis(2015)},  for additional literature. The aim of this 
work is to highlight some recent methodological developments  and  propose some potentially useful research topics. 

Following conventional theory, the standard venue for developing multivariate count time series  models 
requires specification of a joint conditional 
distribution.  Then,  likelihood inference, for a  given  autoregressive  model, provides estimation, testing and all type of standard output. However, 
choosing  a joint count  distribution is a challenging problem.  
There are numerous proposals available in the literature generalizing  univariate Poisson probability mass function (p.m.f); some of these are reviewed in Sec. \ref{seq:distribitions}. 
The main obstacle  is that  the  p.m.f  of a multivariate "Poisson" discrete random vector is often  
of   complicated functional form and therefore maximum likelihood  inference is theoretically and numerically burdensome.
The choice of  joint  distribution for modeling  multivariate count data is an interesting research  topic and some comments will be made throughout this work. 

 The first {modeling}  approach is based on the theory of integer autoregressive  (INAR) models and was initiated by \cite{FrankeandRao(1995)} and \cite{Latour(1997)}. It  was applied  more recently by \citet{PedeliandKarlis(2013b),PedeliandKarlis(2013)}, \citet{Scottoetal(2014)} and  \citet{Darollesetal(2019)}. {INAR models fall within the class of observation-driven models
but because they are defined by means of  thinning operator, (see Def. \ref{def of thin operator} 
and \ref{def of multivariate  thin operator}), they deserve special attention.}
Estimation for INAR  models is based on least
squares methodology and/or likelihood based methods. But, even for  univariate INAR models, likelihood theory is quite
cumbersome, especially when dealing with  higher order autoregressive  models. This methodology is reviewed in Sec. \ref{sec:INAR}.

The second model class  reviewed   is that of parameter driven
models whose dynamics--according to the broad categorization introduced by \citet{Cox(1981)}--
are driven by an unobserved process.   Such models are also called state space models and have found numerous applications; 
see \citet{Zeger(1988)}, \citet{HarveyandFernandes(1989)},
\citet{FahrmeirandTutz(1994)},  \citet{WestandHarrison(1997)}, \citet{DurbinandKoopman(2000)} and 
\citet{Fruhwirth-SchnatterandWagner(2006)}, among others, for contributions on univariate time series  modeling.
Multivariate state space models  were studied by
\citet{Jorgensenetall(1996)} and   \citet{Jungetal(2011)}; 
see also \citet{Ravishankeretal(2014), Ravishankeretal(2015)}, among others, for more recent contributions.
We review these models and we illustrate  that, even though their specification is simple, they still require extensive computational
efforts to be applied  (using either frequentist or Bayesian methods). Developments in this area are reviewed in Sec. \ref{sec:parameter-driven models}.

Section \ref{sec:Observation-Driven Models} goes over observation-driven process. This is the  third class of models included in this work.  Their main characteristic is that    
dynamics  evolve according to past values of the process  plus some noise. For example, ordinary autoregressive models 
belong to this class.  Univariate observation--driven models for count time series have been studied by \citet{ZegerandQaqish(1988)}, \citet{FahrmeirandTutz(1994)},  \citet{RydbergandShephard(2000)}, \citet{KedemandFokianos(2002)},
\citet{Fokianosetal(2009)},  \citet{FokianosandTjostheim(2011)},
\citet{DavisandLiu(2012)},  \citet{AhmadandFrancq(2015)}}, \citet{Doucetal(2017)}, among others.
There is a growing literature within the framework of multivariate observation-driven count time series models;  see \cite{HeinenandRegifo(2007)},
\cite{Liu(2012)}, \cite{Andreassen(2013)}, \citet{Ahmad(2016)}, \citet{Leeetal(2017)}, \citet{CuiandZhu(2018)}, \citet{GourierouxandLu(2019)}, \citet{Fokianosetal(2019)},  \citet{BracherandHeld(2020)}, \citet{Opschooretal(2020)}, \citet{Piancastellietal(2020)}, \citet{ClarkandNixon(2021)} for instance. 
Most of these  studies are  concerned
with linear and/or log-linear  count time series models but other alternatives can be developed.  {Finally, we mention the
work of \citet{DarollesandGourieroux(2015)} who combine  parameter-driven and observation-driven
models  to predict the number of hedge fund defaults as a function of hedge fund
past defaults.}

From a personal perspective,  I think that this research area is still underdeveloped
and there is ample space for exciting new developments. Some more recent works include that of \citet{Veraart(2020)},
who studies  continuous-time models for multivariate count  time series  whose marginal distribution is infinitely
divisible. This construction allows for separate modeling of  serial correlation and the cross-sectional dependence.
Additional work by \cite{Halletal(2019} considers high-dimensional count time series and studies the issues 
of   inference for  autoregressive parameters and the corresponding  network structure by developing a  
sparsity-regularized maximum likelihood estimator. Finally, works by \citet{wu2017}  and  \citet{Dahlhaus(1997)}
are potentially applicable to develop models for high-dimensional and non-stationary data.  It is envisaged that this review  will motivate further research on modeling and inference for multivariate count time series. For instance, simple questions like prediction, diagnostics, testing or development of other more suitable models with/without covariates will require further
studies and theoretical developments. 
The list of references is by no means complete but  further information is given by the therein  and  interested readers should  consult them for further details.

\section{A Review of  Multivariate Count Distributions}
\label{seq:distribitions}

We outline some parametric multivariate count distributions for independent data that have
been found useful for regression analysis. Simple properties of those models are  discussed   
and their connection to  time series data is illustrated  in  Sec. \ref{sec:Observation-Driven Models}. 
The goal is to show that some basic  multivariate distributions are directly applicable 
for fitting regression models and develop inference. There 
are several alternative venues, for instance we can rely  on  copulas  and mixture models; 
see  \citet[Ch. 37]{Johnsonetal(1997)}, \citet[Sec 7.2]{Joe(1997)}  for numerous multivariate count distributions
and \citet[Ch.8]{CameroandTrivedi(2013)} for an in-depth review of multivariate count regression models. 
As a general remark, joint p.m.f.   of  a discrete random vector  often has  
complex functional form  which is far from  being useful to develop likelihood based inference.  
Further recent work and good summaries of up-to date parametric models can be found in \citet{Zhangetal(2017)},
\citet{Inouyeetal(2017)}, \citet{Koochemeskhianetal(2020)}, among others.
 
A multinomial distribution, is traditionally employed for analysis of  multivariate count data by employing a multinomial logistic
regression models.  The multinomial distribution and the case of Dirichlet-multinomial  distribution are not included in this work  (see \citet{Zhangetal(2017)} for more). Those distributions 
are defined by a conditioning argument which might not extend 
to time series context. It is an elementary exercise to show that if $Y_{i}$ is  independent Poisson distributed 
with mean  
$\lambda_{i}$ for $i=1,2,\ldots,d$, then the conditional distribution of $(Y_1, \ldots Y_d)$ given $\sum_{i}Y_{i}=y_{.}$
is multinomial with parameters $y_{.}$ and $\lambda_{i}/ \sum_{j} \lambda_{j}$, for $i=1,2,\ldots,d$. So, the multinomial 
distribution applies to multivariate count modeling but subject to the restriction that it is supported on  
$\{ y= (y_{1}, \ldots y_{d})^T: \sum_{i} y_{i}=y_{.} \}$. This support  constraint  should be considered  cautiously in the context of dependent data. Additionally, this approach provides a conditional likelihood function for estimating regression parameters. The full likelihood function  requires knowledge of the p.m.f. of   $\sum_{i}Y_{i}$.
Similar remark holds for Dirichlet-multinomial regression. 

In what follows, we denote by $Y=(Y_{1}, \ldots, Y_{d})^{T} $ a $d$-dimensional vector of counts whose components are
not necessarily independent.

\subsection{Multivariate Poisson Distributions}
\label{sec:mult Poisson distrtibutions}

This class of distributions  (\citet{KocherlokotaTwo(1992)} and \citet{Johnsonetal(1997)}), 
generalizes the univariate Poisson models. Put $Y_{i}=W_{i}+W$, 
$i=1,2,\ldots,d$ where $W_{i} \sim \mbox{Poisson}(\lambda_{i})$  and 
$W \sim \mbox{Poisson}(\lambda_{0})$ and all $W$'s are independent. Then, the joint p.m.f of $Y$ is given 
\begin{eqnarray}
\mbox{P}[Y=y] =\exp \bigl(-\sum_{i=0}^{d} \lambda_{i} \bigr)  
\biggl( \prod_{i=1}^{d} \frac{ \lambda_i^{y_{i}}}{y_{i}!} \biggr) 
\sum_{k=0}^{\min_{i}y_{i}} \biggl( \prod_{i=1}^{d} {y_{i} \choose k} \biggr) k! 
\biggr( \frac{\lambda_{0}}{\prod_{i=1}^{d} \lambda_{i}} \biggl)^{k}.
\label{eq:mult Poisson distribution}
\end{eqnarray}
The marginals, $Y_{i}$, are Poisson with mean $\lambda_{i}+\lambda_{0}$, for $i=1,2,\dots,d$ and it holds that
$\mbox{Cov}(Y_{i}, Y_{j})= \lambda_{0}$  which is always positive.  In addition,  the parameter $\lambda_{0}$ 
determines all possible pairwise correlations so the resulting model is of limited use.
These facts and the complicated form 
of \eqref{eq:mult Poisson distribution} make this model 
suitable for relatively low dimensional analysis where the marginals $Y_i$ are positively correlated. An E-M type algorithm has been proposed 
by \citet{Karlis(2003)} for  inference but this approach is still hard to implement when the dimension 
$d$ is large. 
Given a vector of regressors, say $X$, an appropriate regression model   $\lambda_i$ as a function of $X$ (in terms of
linear of log-linear link function) and $\lambda_{0}$ is taken as constant . Then using \eqref{eq:mult Poisson distribution} likelihood inference
is straightforward, at least for low dimensions, for more see \citet[Ch. 8.4.1]{CameroandTrivedi(2013)}.

\subsection{Mixed Poisson Models}
\label{subseq:mixed Poisson}

Mixed models provide a general class of multivariate count distributions, see \citet{MarshalandOlkin(1988)}.
{Assume that $Y_1, \ldots, Y_d$ are conditionally independent and Poisson distributed with $\mbox{E}[Y_i] = \lambda_i$, for $i = 1, \ldots,  d$, given $\lambda_1, \ldots, \lambda_d$.}
Suppose that the vector  $\lambda=(\lambda_{1}, \ldots, \lambda_{d})^{T}$  is distributed according to some distribution  $G(\lambda)$. The mixed Poisson 
distribution  is  defined as
\begin{align}
\mbox{P}[Y=y]= \int _{(R^{+})^d} \Bigl[ \prod_{i=1}^{n} \frac{ \exp(-\lambda_i)\lambda_{i}^{y_{i}}}{y_{i}!} \Bigr]  dG(\lambda).
\label{eq:mixed poisson pmf}
\end{align}
Several choices for the  mixing distribution  $G(\cdot)$ exist. But it is always true (provided that appropriate moments exist) that
\begin{equation}
	%\begin{split}
	\mbox{E}[Y]    = \mbox{E}[\lambda],  ~
	\mbox{Var}[Y]  = \mbox{diag}(\mbox{E}(\lambda))+ \mbox{Var}[\lambda]= \mbox{diag}(\mbox{E}(Y))+ \mbox{Var}[\lambda],
	\label{eq:mean and var of mixed Poisson} 
    %\end{split}
\end{equation}
because of the first equality and $\mbox{diag}(x)$ denotes a diagonal matrix whose elements are given by a vector $x$.
A trivial example of \eqref{eq:mixed poisson pmf} is given when $G(.)$ is the Dirac distribution placing its mass at $\lambda$. Then, $Y$ is just a   vector which consists of independent Poisson random variables.  Finite mixtures of multivariate Poisson distributions, with application to clustering,  have been discussed by \citet{KarlisandMeligkotsidou(2007)}.
Another  interesting  case is when  $\lambda$ follows the $d$-dimensional log-normal distribution with parameters $\mu=(\mu_{1}, \ldots, \mu_{d})^{T}$ and $\Sigma=(\sigma_{ij})_{i,j=1,\ldots,d}$. Though no closed formula exists for the p.m.f. of $Y$,  eq. \eqref{eq:mean and var of mixed Poisson} implies that for $i=1,2,\ldots,d$
\begin{eqnarray*}
	\mbox{E}[Y_{i}] &= & \exp(\mu_{i}+ 0.5\sigma_{ii}), \\
	\mbox{Var}[Y_i] & = & \mbox{E}[Y_{i}]+ \mbox{E}^{2}[Y_{i}] \bigl( \exp(\sigma_{ii})-1\bigr), \\
    \mbox{Cov}(Y_{i}, Y_{j}) &= & \mbox{E}[Y_{i}] \mbox{E}[Y_{j}]\bigl( \exp(\sigma_{ij})-1\bigr).
\end{eqnarray*}
In general, denote by $y_{(k)}=y(y-1)....(y-k+1)$, the so called  falling factorial. Then \eqref{eq:mixed poisson pmf}
shows that multivariate factorial moments are computed by using  simple properties of  Poisson distribution, i.e.
\begin{align*} \displaystyle 
	\mbox{E}\Big[ \prod_{i=1}^{d} Y_{i, (r_{i})} \Big] & = \mbox{E} \Big\{ \prod_{i=1}^{d} \mbox{E}\Big[Y_{i,(r_{i})} \mid \lambda_{i}  \Big] \Big \} = \mbox{E} \Big[  \prod_{i=1}^{d} \lambda^{r_{i}}_{i}  \Big], 
\end{align*} 
Furthermore, following \citet{JohnsonandKotz(1992)} joint moments are given by 
\begin{align*}
		\mbox{E}\Bigl[ \prod_{i=1}^{d} Y^{r_{i}}_{i} \Bigr] & = \sum_{l_{1}}^{r_{1}} \ldots \sum_{l_{d}=0}^{r_{d}} \prod_{i=1}^{d} s(r_{i}, l_{i}) 
		\mbox{E} \Bigl[  \prod_{i=1}^{d} \lambda^{l_{i}}_{i}  \Bigr],
\end{align*}
where $s(r,l)$ are the Stirling numbers of the second kind. 

Next, it is shown that the multivariate negative-multinomial distribution is recovered by means of \eqref{eq:mixed poisson pmf}. This is well-known in the univariate case. A multivariate  negative-multinomial distribution has 
p.m.f. which is  given by
\begin{equation}
	\mbox{P}[Y=y] = \frac{ (r+ \sum_{i=1}^{d} y_{i})!}{ (\prod_{i=1}^{d}y_{i}!) (n-1)!} p_{0}^{r} 
	\prod_{j=1}^{d}  p^{y_{j}}_{j},
\label{eq:pmf of negative multinomial}
\end{equation}
where $y_{j}=0,1,2, \ldots$ for  $j=1,\ldots,d$, $r >0$ and $0 < p_{j} <1$, $j=0,1,\ldots d$ satisfying 
$p_{0}=1-\sum_{j=1}^{d} p_{j}$. The parameters $p_{j}$, $j=1,2,\ldots,d$ denote the probabilities of obtaining different failures and thus $p_{0}$ is the probability of success in an experiment terminating to $r$ failures; see
\citet[Ch. 36]{Johnsonetal(1997)} and \citet[Ch. 7.2]{Joe(1997)} for more. Note that   $r$ might assume real values in applications;  it is  short of "dispersion" parameter though the concept of dispersion--that is when the variance exceeds the mean--is quite vague in the multivariate case we consider. In this case, it can be shown that all 
pairwise correlations  between the components of $Y$ are positive.

It is shown next that \eqref{eq:pmf of negative multinomial} is obtained as
mixed Poisson model by using \eqref{eq:mixed poisson pmf} 
assuming that, conditionally on a Gamma distributed random variable $\theta$,
say  $\theta \sim \mbox{Gamma}(\beta, \beta)$,  $Y_{j}$ is conditionally Poisson distributed with mean  $\lambda_{j} \theta$, $j\in\{1,\dots,d\}$.  Then 
\begin{align}
	\mbox{P}[Y=y] & =  \int_{0}^{\infty}  \mbox{P}[Y=y \mid \theta]dG(\theta) \nonumber \\ 
	              & =  \frac{ \Gamma( \beta + \sum_{i=1}^{d}y_{i})}{ (\prod_{i=1}^{d}y_{i}!) \Gamma(\beta)}
	              \Bigl(  \frac{\beta}{\beta+\sum_{i=1}^{d} \lambda_{i}}  \Bigr)^{\beta} 
	              \prod_{j=1}^{d}
                  \Bigl( \frac{\lambda_{j}}{\beta+\sum_{i=1}^{d} \lambda_{i}}   \Bigr)^{y_{j}}, 	              
\label{eq:mixed Poisson negmult pmf}
\end{align}
where $\Gamma(\cdot)$ denotes the Gamma function. Obviously \eqref{eq:pmf of negative multinomial} holds. 
Clearly, this model implies   that the random variable  $\theta$ accommodates  common unobserved heterogeneity; see \citet{MunkinandTrivedi} for
simulated maximum likelihood estimation for this particular class.

Recall \eqref{eq:pmf of negative multinomial}. Then, given a covariate vector $X$, a multinomial logistic 
regression model (see \citet{Agresti(2002)}) is employed to link $X$ with the probabilities $p_{j}$, $j=1,2,\ldots, (d+1))$.
Furthermore, the model can be extended to include a log-linear model for $r$ (which can be positive real in general, see the previous case)--for more 
details see \citet{Zhangetal(2017)}.

\subsection{Copula approaches}
\label{subseq:Copulamodels}

Copula-based construction of multivariate count distributions is an active topic of research; see  \citet{Nikoloulopoulos(2013b)} and \citet{Inouyeetal(2017)}
for nice surveys.  Copulas are useful  because of  Sklar's theorem (\citet{Sclar(1959)}) which shows that marginal distributions  are  combined to give a joint distribution when applying  a 
copula, i.e. a $d$-dimensional distribution function all of whose  marginals are standard uniforms; the book by \citet{Nelsen(1999)} gives  a thorough  introduction to copulas. 

Even though copulas provide an appealing  methodology for constructing joint distribution functions they pose  
challenging  issues when used for  discrete data analysis. 
First, the presence of ties in count data (several zeroes, for example)
makes them non-identifiable--\citet{GenestandNeslehova(2007)},  in particular pp. 507-508--illustrate the lack of identifiability.  An additional issue is that of the  likelihood  function's computationally difficulty   for estimating  unknown parameters. For a discrete random vector, whose cumulative distribution function (c.d.f)  is $F$, its p.m.f. involves $2^{d}$ finite differences of $F$, i.e. 
\begin{equation*}
	\mbox{P}[Y=y]= \sum_{l_{1}=0,1} \ldots \sum_{l_{d}=0,1} (-1)^{l_{1}+\ldots+l_{d}}~
	\mbox{P}[Y_{1} \leq y_{1}-l_{1}, \cdots, Y_{d} \leq y_{d}-l_{d}].
\end{equation*}
Computational methods 
using  Bayesian data augmentation have  been developed by \citet{SmithandKhaled(2012)}, among others;  the survey of  \citet{Smith(2013)} provides  references on Bayesian methodology for discrete data copula modeling.

In the rest of this section I  describe  a different approach for employing  copulas to model multivariate count data.
The methodology is based on the work by   \citet{Fokianosetal(2019)} who advanced  a particular  data generating process for multivariate
count time series analysis; this topic is discussed in Sec. \ref{sec:Observation-Driven Models}.  
Initially,  the idea is illustrated  for  the case of i.i.d random vectors. The intent  is to  introduce a data generating process which keeps \emph{all}  marginal distributions of the vector $Y$ to be  
Poisson distributed   and, at the same time, it  allows
for arbitrary dependence among them. This is accomplished  by  appealing to elementary   properties of  Poisson process.
An explicit account of this construction is given by the  following algorithm--recall that $Y_{i}$, $i=1,2,\ldots,d$ is the $i$'th-component 
of the count vector $Y$ whose mean is $\lambda_{i}$. 

\begin{enumerate}
	\item Let ${ U}^{l}=(U_{1 ,l}, \ldots, U_{d,l})$ for $l=1,2,\ldots, K$, be a sample from a $d$-dimensional
	copula $C(u_{1}, \ldots, u_{d})$. Then $U_{i,l}$, $l=1,2,\ldots, K$ follow marginally the uniform distribution on $(0,1)$, for
	$i=1,2,\ldots,d$.
	\item Consider  the transformation
	%\begin{eqnarray*}
	$X_{i, l} = -{ \log U_{i, l}}/{\lambda_{i}}, ~i=1,2, \ldots, d.$
	%\end{eqnarray*}
	Then, the marginal distribution of $X_{i,l}$, $l=1,2,\ldots, K$
	is exponential with parameter $\lambda_{i}$, $i=1,2,\ldots,d$.
	\item
	{If $X_{i,1} >1$ set $Y_i=0$, otherwise 
	$Y_{i} = \max \left\{K:~ \sum_{l=1}^{K} X_{i, l} \leq  1 \right\}, ~i=1,2,\ldots,d.$}
	%\end{eqnarray*}
	Then ${Y}=(Y_{1}, \ldots, Y_{d})^{T}$ is marginally  realization   of a    Poisson process with parameter ${\lambda}$.
	\item Repeat steps 1-3  $n$ times  to generate a sample $Y^{j}$, $j=1,2,\ldots,n$. 
\end{enumerate}

The algorithm generates i.i.d. random vectors whose  dependence among their components is introduced by a
copula structure on the waiting times of the Poisson process. In other words,  the copula is imposed on the uniform random
variables generating the exponential waiting times. The end result gives a sample of multivariate discrete random variables with
Poisson marginals. 
%{The expected complexity of this algorithm grows linearly in $d (\max_{i} \lambda_{i})$ since the 
%generation of marginal Poisson random variables has  expected order complexity $\lambda_{i}$, $i=1,2, \ldots,d$, (see \citet[Ch.10]{Devroye(1986)})
%and generation of copula realizations  requires  $(d+1)$ uniform random variables; see \citet{MarshalandOlkin(1988)} and 
%the R package \texttt{copula} by \cite{CopulaR}. There exist more efficient algorithms for generating Poisson random variables (see
%\citet[Ch.10]{Devroye(1986)}) but because the focus in on time series modeling, this algorithm is preferred because
%it does not alter standard Poisson process properties marginally. }
 The methodology  can be extended to other discrete  marginal distributions  provided that  
they can be generated by continuous inter arrival times.
For instance, suppose that $Y_{i}$ is marginally mixed Poisson with mean $\theta  \lambda_{i}$ where $ \theta$ 
satisfies $\mbox{E}[\theta ]=1$.  Many families of count distributions,
including the negative binomial, can be generated by this construction; see Sec. \ref{subseq:mixed Poisson}.  Then steps
1-4 of the above algorithm  are still useful for generating  multivariate random vectors whose marginals are not necessarily Poisson.
Indeed, generating at the first step an additional random variable  $\theta$, 
define again at step 2 the waiting times by $X_{i, l} = -{ \log U_{i, l}}/{ \theta \lambda_{i}}, ~~~i=1,2, \ldots, d.$ Then, the distribution of
$X_{i}$ is mixed exponential and therefore steps 3-4 deliver a realization of a count vector whose marginal distribution is mixed Poisson.

\begin{figure}[h]
	\centering
	\includegraphics[width=0.8\linewidth, height=0.4\textheight]{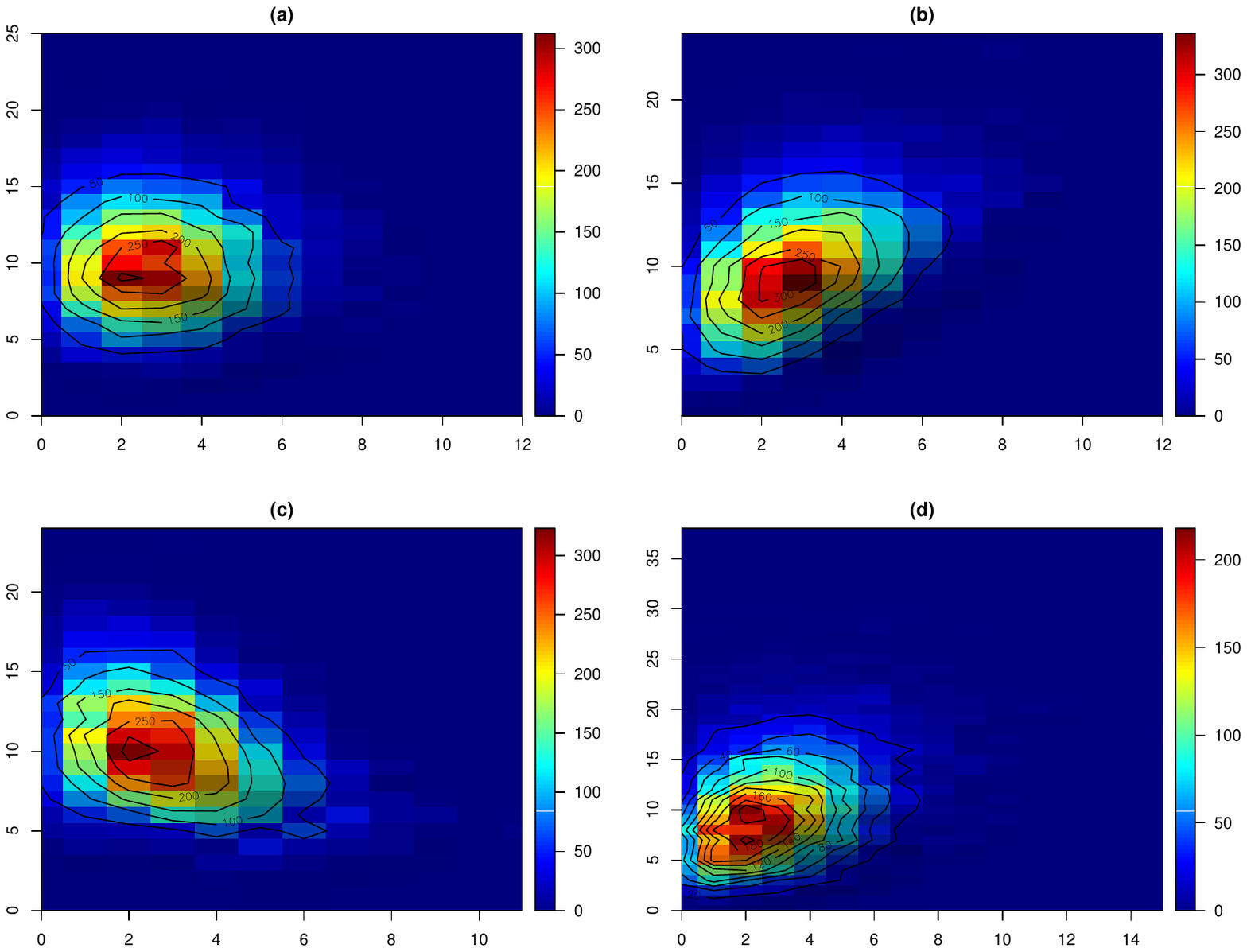}
	\caption{Joint p.m.f of a bivariate count distribution using the copula construction as outlined by steps 1--4. Results are based on a Gaussian copula  with correlation coefficient $\rho$. (a) $\rho=0$ (independence) (b) $\rho=0.8$ (positive correlation) (c) $\rho=-0.8$ (negative correlation). Plots are based on 10000 independent observations  where the marginals are Poisson with $\lambda_{1}=3$ and $\lambda_{2}=10$.  (d) Joint p.m.f of negative multinomial distribution \eqref{eq:mixed Poisson negmult pmf}. Results are based on 10000 independent observations  with $\beta=10$ and  $\lambda_{1}=3$ and $\lambda_{2}=10$.  }
	\label{fig:distplot}
\end{figure}

The joint p.m.f. of $Y$, based on the above construction, is shown in  Fig. \ref{fig:distplot}.
Plots (a), (b) and (c) show the case of independence, positive and negative correlation, respectively, when using  
a Gaussian copula. The algorithm delivers desired marginals and, in addition, shows that all type of different correlations
can be achieved. Fig. \ref{fig:distplot}(d) shows that joint p.m.f. of a  negative multinomial vector, see \eqref{eq:mixed Poisson negmult pmf} with the same $\lambda$ parameters values  used for the previous algorithm and $\beta=10$. The positive correlation between the vector components is obvious. 

The algorithm, when implemented, requires clear distinction between the resulting copula to the 
vector of counts and the copula imposed on waiting times. The transformation from waiting times to counts is stochastic, and while the copula as such is invariant to one-to-one deterministic transformations, we do not have such a transformation in this case. Hence,  
instantaneous correlation among
the components of the count vector  is not equal to the correlation induced by the copula imposed to the vector of waiting times.  \emph{Interpretation of the instantaneous correlation found in data  is related to the correlation of the vector
of waiting times and should be done with care.}  {Recalling Fig. \ref{fig:distplot} and calculating 
the sample correlation coefficient, it is found  that it is equal to 0.34 (-0.34) when the data are generated with copula parameter 
$\rho >0$ ($\rho <0$), respectively.}

This approach is different from the methodology advanced  by  \citet{DenuitandLambert(2005)} who  
employ   the continued extension. Those authors  add   noise of the form   $U-1$, where $U$ is
standard uniform, to  counts so that those  are transformed into   continuous random variables. In doing so, the copula identifiability problem of   is bypassed. An analogous approach, based on the distributional transform 
which adds a random jump to the c.d.f. of the discrete variable,
has been studied by \citet{Ruschendorf2013}. An interesting decomposition of the joint p.m.f. of a discrete 
random vector has been discussed by \citet{Panagiotelisetal(2012)} using the idea of pair-copula construction (\citet{Czasdo(2010)}) by utilizing the concept D-vine copulas (\citet{BedfordandCooke(2001), BedfordaandCooke(2002)}). 

Further work on copulas, in the context of generalized linear models, is given by \citet{Song(2000)} and  \citet{Songetal(2009)}. The latter reference   employs  Gaussian copulas, for multivariate 
regression analysis of continuous, discrete, and mixed correlated outcomes under the generalized linear models (GLM)
framework (\citet{NelderandWedderburn(1972)} and \citet{McCullaghandNelder(1989)}).  More recently, 
\citet{Yangetal(2020)}  consider discrete regression models and copula estimation arguing that 
inclusion of continuous covariates implies consistent estimation of the unknown copula. In addition,
\citet{Jiaetal(2020)} employ a latent Gaussian process and a distributional transformation to construct stationary
univariate count time  series models  with  flexible correlation features such that their  marginal distribution can be prespecified. 

%Their methodology includes a
%non parametric estimator of copulas which additionally helps on identifying a parametric copula form.

\subsection{Additional models}

There are additional approaches for defining a multivariate count distribution; \citet[Ch.9]{Joe(1997)} and  \citet{Inouyeetal(2017)} review construction 
of multidimensional  Poisson p.m.f 
by appealing to full conditional distributions and Markov random fields; see \citet{Besag(1974)}
for the so called auto-Poisson model. Additional models include the Sarmanov  and bivariate Hurdle distributions among others; see \citet[Ch. 8]{CameroandTrivedi(2013)}. These models are mentioned for completeness of presentation but their properties have not been fully explored in the literature, to the best of my knowledge.

\section{Integer AR models}
\label{sec:INAR}

Integer Autoregressive (INAR) models   deserve
special consideration due to the {\em thinning} operation.
The calculus of thinning operators provides useful insight into
the probabilistic properties of those  processes by employing  the
simple device of summing up a  \emph{random number} of integer-valued random variables 
(see  \citet{SteutelandHarn(1979)}). The case that has attracted more attention is when
the summands consist of  an independent and identically distributed (iid) sequence of Bernoulli random variables.

\subsection{The thinning operator}

Define
the generalized Steutel and van Harn operator (see \citet[Def. 1.1]{Latour(1997)}) as follows:

\begin{Definition}
\rm Suppose that $X$ is a non--negative integer random variable.  The generalized thinning  operator, denoted by $\circ$, is defined as
$$
\alpha \circ
X
=
\left\{
  \begin{array}{ll}
     \sum_{k=1}^{X} I_{k}, & \hbox{$X >0$;} \\
      0, & \hbox{$X=0$.}
  \end{array}
\right.
$$
where $\left\{ I_{k}, k \in \mathbb{N} \right\}$ is  a sequence of iid integer  random
variables--independent of $X$--with mean $\alpha$ and variance $\beta$.
\label{def of thin operator}
\end{Definition}

The sequence
$\left\{ I_{k},  k \in \mathbb{N}  \right\}$ is called  counting series.
If $\left\{ I_{k} \right\}$ is an iid sequence of Bernoulli random variables, then
$\alpha \circ X$ counts the number of
successes in a random number of Bernoulli trials where the probability
of success $\alpha$  remains constant throughout the experiment so that
given $X$, $ \alpha \circ X$ is a binomial random variable with
parameters $X$ and $\alpha$. {In this case, we call the thinning operator as binomial thinning 
operator. General thinning operators are discussed  by \citet[Sec.2]{Davisetal(2021)}
and \cite{Joe(2016)}, among others.}

 Numerous properties can be proved for the thinning operator, for instance
it can be shown that $\mbox{E} \left[ \alpha \circ X \mid X
\right]  =   \alpha X$,  $ \mbox{E} \left[ \alpha \circ X \right]  =   \alpha \mbox{E}\left[ X \right]$,  $ \mbox{Var}
\left[ \alpha \circ X \mid X \right]  =  \beta  X$ and
$\mbox{Var} \left[ \alpha \circ X  \right]  = \alpha^{2} \mbox{Var}\left[ X
\right] + \beta \mbox{E}\left[ X \right]$, provided that appropriate moments of $X$ exist.

Definition \ref{def of thin operator} can be extended to a {non-negative} integer $d$-dimensional random vectors. For $i,j=1,2,\ldots, d$,
define $\left\{ I_{ij;k},  k \in \mathbb{N}  \right\}$ an array of counting series
such that $\mbox{E}[I_{ij}]=\alpha_{ij}$ and $\mbox{Var}[I_{ij}]=\beta_{ij}$.  Let $A=(\alpha_{ij})$, $B=(\beta_{ij})$ be the corresponding $d \times d$ matrices. Then, Definition \ref{def of thin operator} can be extended as follows:

\begin{Definition}
\rm Suppose that $X=(X_{1}, X_{2}, \ldots, X_{d})^{T}$ is $d$-dimensional integer-valued random vector with all  components being  non-negative and
denote by $A \circ=(\alpha_{ij} \circ)$ a $d \times d$ matrix of thinning operators  whose each element is
given  by Def.  \ref{def of thin operator} with  corresponding array of counting series $\left\{ I_{ij;k},  k \in \mathbb{N}  \right\}$. Then the\emph{ multivariate thinning} operator is defined as
\begin{align*}
  A \circ X  &  = \left(
                  \begin{array}{c}
                    \sum_{j=1}^{d} \alpha_{1j} \circ X_{j} \\
                     \vdots \\
                    \sum_{j=1}^{d} \alpha_{dj} \circ X_{j}  \\
                  \end{array}
                \right).
\end{align*}
\label{def of multivariate  thin operator}
\end{Definition}
Similar to the univariate case, it can be shown that $\mbox{E}[A \circ X]=A \mbox{E}[X]$ and $\mbox{Var}[A \circ X] = 
\mbox{diag}(B E[X])+ A \mbox{Cov}[X]A^{T}$.

\subsection{The multivariate  INAR model}

The multivariate thinning operator serves as basic tool to develop multidimensional  integer AR models (abbreviated by MINAR) of order $p$. A $d$-dimensional time series $\{Y_{t}, t \in \mathbb{Z}\}$ is
called multivariate INAR($p$) process if it satisfies 
\begin{eqnarray}
\label{eq:definition of INAR(p) model}
  Y_t &= \sum_{i=1}^{p} A_{i} \circ Y_{t-i} + \epsilon_{t},
\end{eqnarray}
where $\{ \epsilon_{t}, t \in \mathbb{Z}\}$ is a sequence of iid integer-valued random  vectors with mean $\mbox{E}[\epsilon_{t}]=\mu_{\epsilon}$ and $\mbox{Var}[\epsilon_{t}]=\Sigma_{\epsilon}$ which  is 
independent of all thinning operators $A_{i}$, $i=1,2,\ldots, p$ and $A_{p} \neq 0$. 
Denote by  $I_{d}$ the $d$-dimensional identity matrix. 
Provided that the roots of polynomial $\det(I_{d}-A_{1}z-\cdots A_{p} z^{p})$  are all located  outside the unit circle,  then \citet[Prop. 3.1]{Latour(1997)}
shows that there exists an almost surely unique integer-valued strictly stationary process that satisfies \eqref{eq:definition of INAR(p) model} and such that $\epsilon_{t}$ is independent 
of $Y_{s}, ~ s < t$.  For the case of $d=1$, this condition is equivalent to $0 < \sum_{i=1}^{p} A_{i} <1 $, see \citet{Duandli(1991)}. The univariate INAR($p$) processes have been introduced by \citet{OshandAlzaid(1987)},  \citet{AlzaidandOsh(1990)}.  In the same vein, multivariate {INAR  moving average  (MINARMA) models can be  defined but 
they will  not be discussed any further.}

We assume  the stability condition for MINAR($p$) processes holds true.
Recall \eqref{eq:definition of INAR(p) model} and consider  the special case of $p=1$. By taking expectations in both sides of \eqref{eq:definition of INAR(p) model}, $\mbox{E}[Y_{t}]=(I_{d}-A_{1})^{-1}\mu_{\epsilon}$. Multiplying both sides of \eqref{eq:definition of INAR(p) model} by $Y_{t+h}$ and taking expectations, it follows that 
$\mbox{Cov}(Y_{t}, Y_{t+h})= A_{1}^{h} \mbox{Var}(Y_{t}) $, where $\mbox{Var}(Y_{t})= A_{1} \mbox{Var}(Y_{t})A_{1}^{T}+ \mbox{diag}(B \mbox{E}[Y_{t}])+ \Sigma_{\epsilon}$ for $h=0, \pm 1, \pm 2 \cdots$.  As a final remark, \citet[Prop. 4.1]{Latour(1997)} shows that a MINAR($p$), which satisfies the
stability condition discussed earlier, has identical second order properties with an ordinary vector AR($p$) (VAR) model (see \citet{Tsay(2014)}).  Consider again the case $p=1$ for  \eqref{eq:definition of INAR(p) model}. Then $\{Y_t\}$ is 
represented by a  VAR(1) process of the form 
\begin{eqnarray}
	Y_{t}=  \omega+ A_{1} Y_{t-1} + \zeta_{t}
\label{eq:VAR(1) for INAR(1) representation}
\end{eqnarray}
where $\{ \zeta_t\}$ is a  white noise process with covariance matrix $\Sigma_{\zeta}= 
\mbox{diag}(B \mbox{E}[Y_{t}])+ \Sigma_{\epsilon}$ and $\omega=\mbox{E}[Y_{t}]$. 
This fact has important consequences on estimation. For instance least squares estimators (LSE)
directly applies to this class of  models but subject to restriction that all unknown  coefficients are positive.

\subsection{Estimation}
\label{sec:estimation}

Besides LSE, likelihood estimation has been also developed for estimating the unknown parameters of the model
\eqref{eq:definition of INAR(p) model}. Both methods are discussed next by assuming that $Y_{1}, \ldots, Y_{n}$ is 
a sample from a  MINAR(1) model--this is done mostly for convenience. 

LSE are  computed and studied by using \eqref{eq:VAR(1) for INAR(1) representation}. Let 
$$
\underbrace{Y}_{(n-1)\times d} = \underbrace{X^{\star}}_{(n-1)\times (d+1)} ~ \underbrace{\beta}_{(d+1) \times d} + 
\underbrace{Z}_{(n-1) \times d}
$$
where the $i$'th row of  $Y$, $X^{\star}$ and $Z$  is given by  $Y_{i+1}^{T}$,  $(1, Y^{T}_{i})$ and  $\zeta^{T}_{i}$, $i=1, 2,\ldots, (n-1)$, respectively. The
regression matrix  parameter is denoted by $\beta$, i.e.  $\beta=(\omega, A_{1})^{T}$.  Then the LSE of $\beta$ is denoted by $\widetilde{\beta}$
and is equal to
\begin{align}
\begin{aligned}
	\widetilde{\beta} & = \biggl( X^{\star T}X^{\star} \biggr)^{-1} X^{\star T} Y 
	                   =	\biggl(
	\sum_{t=2}^{n} 
	\begin{bmatrix}
		1  & Y^{T}_{t} \\
		Y_{t} & Y_{t} Y^{T}_{t} 
	\end{bmatrix} 
		\biggr)^{-1} 
		\biggl( 
		\sum_{t=2}^{n} 
	\begin{bmatrix}
		Y^{T}_{t} \\
	    Y_{t-1}Y^{T}_{t}
	\end{bmatrix}
		\biggr). 
\label{eq:LSE for MINAR(1)}
\end{aligned}
\end{align}
Based on this define the residual matrix $\widetilde{Z}=Y-X^{\star} \widetilde{\beta}$ to obtain an estimator
of $\Sigma_{\zeta}$  by 
\begin{align*}
	\widetilde{\Sigma}_{\zeta}= \frac{1}{n-d-2}  \widetilde{Z} \widetilde{Z}^{T},
\end{align*}
where the numerator $n-d-2=(n-1)-(d+1)$ is equal to the effective degrees of freedom minus the number
of parameters estimated for each component series. More precisely, it can be shown that 
$\mbox{E}[\widetilde{\beta}]= \beta$ and $\mbox{E}[\widetilde{\Sigma}_{\zeta}]= \Sigma_{\zeta}$.
Moreover, as $n \rightarrow \infty$,  and assuming suitable regularity conditions are fulfilled
\begin{align*}
	\sqrt{n}  \vect \bigl( \widetilde{\beta}-\beta \bigr) \stackrel{D}{\Longrightarrow} N(0, \Sigma_{\zeta} \otimes H^{-1}),
\end{align*}
where $\vect(.)$ denotes the vec operator, $\otimes$ is the Kronecker product and the $(d+1) \times (d+1)$ matrix $H$ is the limit (in probability) of  $X^{\star T}X^{\star}/n$. A rigorous statement is developed along the lines of \citet{Latour(1997)} and \citet[Lemma 3.1]{Lutkepohl(2005)}.  This result holds when the true parameters belong to the interior of parameter space employing 
unconstrained  optimization.  The asymptotic distribution of  LSE,  under the constraint  that all
elements of the matrix $A_{1}$ are positive, is  an open problem.  

Besides LSE, conditional likelihood estimation is developed (recall again \eqref{eq:definition of INAR(p) model} with $p=1$) by maximizing   the likelihood function 
\begin{align*}
	L(\beta)=\prod_{t=1}^{n} \mbox{P}_{\beta}[Y_t=y_{t} \mid Y_{t-1}=y_{t-1}],
\end{align*}
when  imposing a multivariate distribution on the error term $\epsilon_{t}$. In general 
the conditional transition  is given by the $d$-dimensional convolution 
\begin{align*}
\mbox{P}_{\beta}[Y_t=y_{t} \mid Y_{t-1}=y_{t-1}] = \sum_{k=0}^{y_{t}} P_{\beta}[A_{1} \circ Y_{t-1}=y_{t}-k] P[\epsilon_{t}=k].
\end{align*}	
Both the above equation implies that the log-likelihood function is given by 
\begin{align*}
	l(\beta) \equiv \log L(\beta)= \sum_{t=1}^{n} \log\Bigl(\sum_{k=0}^{y_{t}} P_{\beta}[A_{1} \circ Y_{t-1}=y_{t}-k] P[\epsilon_{t}=k] \Bigr).
\end{align*}
This short discussion shows that  the task of computing the log-likelihood function is daunting  even in the simple case $p=1$.  
Some simplifications occur when we assume that the matrix $A_{1}$ is diagonal and by applying
pairwise likelihood methodology as in \citet{PedeliandKarlis(2013b)}. Such an approach can deliver some insights for data analysis but it will be difficult to be justified for multivariate dynamic systems. 
Generally, likelihood methods are not 
suitable for this class of models because of their complicated structure. The problem  complexity  increases 
when both $p$ and $d$ grow but sparsity ideas  (see \citet{Hastieetal(2015)}) will be helpful.

\subsection{Prediction}

Consider again the case of model  \eqref{eq:definition of INAR(p) model} for $p=1$. We briefly discuss prediction for
this particular case to overcome cumbersome notation. Similar to the case of VAR($p$) models, a  MINAR($p$) model is  
written  as a "big" MINAR(1) so these results suffice to develop a general point of view. 

The one-step ahead predictor of the MINAR(1) process is easily calculated by
\begin{align*}
	\mbox{E}[Y_{t+1} \mid Y_{t}] =\mbox{E} \left[  A_{1} \circ Y_{t}+ \epsilon_{t} \mid Y_{t} \right]=A_{1} Y_{t}+\mu_{\epsilon}.
\end{align*}
Therefore, by recursion
\begin{align*}
	\mbox{E}[Y_{t+h} \mid Y_{t}] = A_{1}^{h}Y_{t} + \bigl( I_{d}+A_{1}+ \cdots+A_{1}^{h-1} \bigr) \mu_{\epsilon}. 
\end{align*}
Define $V(h)=\mbox{Var}[ Y_{t+h} \mid Y_{t} ]$. Then 
\begin{align*}
	V(1) = \mbox{Var} \bigl( A_{1} \circ Y_{t}+ \epsilon_{t} \mid Y_{t} \bigr) = \mbox{diag}(B \mbox{E}[Y_{t}])+ \Sigma_{\epsilon}.
\end{align*}
The law of total variance shows that 
\begin{align*}
	V(h) & = \mbox{E} \Bigl[   \mbox{Var}  \bigl(Y_{t+h} \mid Y_{t+1}  \bigr) \mid Y_{t} \Bigr]  
	       +  \mbox{Var} \Bigl[  \mbox{E} \bigl( Y_{t+h} \mid Y_{t+1} \bigr) \mid  Y_{t} \Bigr] \\
	     & = \mbox{E}[ V(h-1) ] + A^{h-1}_{1} V(1) ( A^{h-1}_{1})^{T}
\end{align*}
Provided that  $A_{1}$ is diagonal, some  simplifications of the previous formulas have been   proved by \cite{PedeliandKarlis(2013)}. The complexity of prediction problem increases as  both dimension and  model order increase. In addition, care should be taken when  unknown parameters are replaced by their estimators.  

Closing this section, we  mention that properties of multivariate INAR  models are well understood for  low-dimensional
data and under tangible  assumptions. It is interesting to consider properties of the multivariate thinning operator
in high-dimensions and investigate the problem of estimation and prediction.

\section{Parameter-Driven Models}
\label{sec:parameter-driven models}

A parameter driven model, according to classification introduced by \citet{Cox(1981)}, is a time series driven by an unobserved process (as opposed to past process values; see Sec \ref{sec:Observation-Driven Models}).   For multivariate count series, state-space models were studied in \citet{Jorgensenetall(1999)} and \citet{Jungetal(2011)}; see \citet{Ravishankeretal(2014)}, and \citet{Ravishankeretal(2015)}, among others, for more recent contributions. The approaches that have been taken for estimation are based either on  likelihood or full Bayesian methods. We review some of these works. 

The model proposed by   \cite{Jorgensenetall(1999)}, is closely related to the theory of mixed Poisson distributions, see 
Sec. \ref{subseq:mixed Poisson}. It 
assumes that  the conditional distribution of the $i$'th component of the 
multivariate count series at time $t$,  $Y_{i,t}$, given an unobserved univariate time-varying process
\footnote{For this part of the text, I replace the notation $\lambda_t$ by $\theta_t$ because it denotes 
a univariate time-varying mean process.}  $\theta_{t}$, is Poisson
distributed  with mean $a_{i,t}\theta_{t}$  such  that $a_{i,t}=\exp({c}_{t}^{T}\mathbf{\alpha}_{i})$,
where  ${c}_t$ are  $k$-dimensional time-varying covariate vectors and  ${\alpha}_{i}$ are $k$-dimensional regression parameters, $i=1, \ldots, d$.   Assume that $\theta_{0}=1$ and the conditional distribution of  $\theta_{t}$ given $\theta_{t-1}$ is Gamma  with mean $b_{t} \theta_{t-1}$ and a squared coefficient of variation of form $\sigma^{2}/\theta_{t-1}$. The parameters $b_{t}$ depend on the so called long--term covariates ${z}_{t}$ through $b_{t}=\exp(\Delta {z}_{t}^{T}\beta)$, where $\Delta{z}_{t} ={z}_{t}-{z}_{t-1}$ and ${z}_{0}=0$, $\sigma^{2}$ denotes a dispersion parameter and {$\beta$ is regression coefficient}.  It can be shown that 
\begin{align*}
	\mbox{E}[\theta_t] & =b_{1}\ldots b_{t} 
\end{align*}
which implies that  $\log( \mbox{E} [ \theta_{t} ]) ={z}_{t}^{\prime} \beta$. In addition, for $h \geq 0$,  
\begin{align*}
\mbox{Var}( \theta_{t} ) = \phi_{t} \mbox{E}[ \theta_{t} ] \sigma^{2}, ~~
\mbox{Cov}( \theta_{t}, \theta_{t+h} ) & = \phi_{t} \mbox{E}[ \theta_{t+h} ] \sigma^{2},
\end{align*}
where  $\phi_{t}=b_{t}+b_{t}b_{t-1}+b_{t}b_{t-1}\ldots b_{1}$. Set 
${a}_{t}=(a_{1,t}, \ldots, a_{d,t})^{T}$ and $A_{t} = \mbox{Diag}(a_{1,t}, \ldots , a_{d,t})$. Then (compare with 
\eqref{eq:mean and var of mixed Poisson} when $a_{i,t}=1$ for all $i$ and $t$)
\begin{align*}
\mbox{E} [ {Y}_{t}]  = {a}_{t} \mbox{ E} [ \theta_{t} ], ~~
\mbox{Var}({Y}_{t})  = A_{t} \mbox{E}[ \theta_{t} ]+ {a}_{t} {a}_{t}^{T} \phi_{t} \sigma^{2} \mbox{E}[ \theta_{t} ].
\end{align*}
This last result  shows that the variance matrix of ${Y}_{t}$ consists of two components; (a) a Poisson variance and (b) a
type of  "overdispersion" component. The authors discuss Kalman prediction and filtering, for the  log-linear model $\mbox{E}[Y_{i,t}]= \exp({c}_{t}^{T} {\alpha}_{i} + {z}_{t}^{T}\beta)$, relying on previous calculations.

The above model is closely related to the model of \citet{Jungetal(2011)} which in turn generalizes that of 
\citet{Wedeletal(2003)} who developed a comprehensive class of factor  models for  multivariate truncated count data.
The former authors assume that $Y_{i,t}$,  conditionally on $\lambda_{i,t}$, are independent  Poisson distributed random variables with mean $\lambda_{i, t}$, $i=1,2,\ldots,d$ and for all $t$. By considering the $d$-dimensional 
time-varying vector  process $\lambda_{t}=(\lambda_{1,t}, \ldots, \lambda_{d,t})^{T}$, it is assumed that 
\begin{align*}
	\log \lambda_{t} = \omega+ \Gamma f_{t},
\end{align*} 
where $\omega$ is $d$-dimensional vector of parameters, $\Gamma$ is a $d \times s$ matrix of factor loadings, $f_{t}$
is an $s$-dimensional vector of latent random factors and the $\log(.)$ function is taken componentwise. 
Further, the components of $f_t$ are decomposed to similar  subsets which are assumed to follow independently Gaussian AR(1)
model.   \citet{Jungetal(2011)} develop estimation under Poisson and Negative binomial distribution by employing efficient 
importance sampling and apply this methodology  to  numbers of trades, in 5-min intervals, for five New York Stock Exchange stocks from two industrial sectors. 

Similarly, \citet{WangandWang(2018)} assume that $\mbox{E}[Y_{i,t}]= \mbox{E}[\lambda_{i,t}] \varepsilon_{i,t}$,
where $\varepsilon_{i,t}$ is the $i$'th component of a $d$-dimensional hidden process $\varepsilon_t$ such that 
$\mbox{E}[\varepsilon_{i,t}]=1$ (see also  \citet{Zeger(1988)}, \citet{DavisandWu(2009)} and  \citet{ChristouandFokianos(2014)}). This assumption implies that 
\begin{align*}
	\mbox{Cov}[Y_{i,t}, Y_{j,u}] = \mbox{E}[\lambda_{i,t}] \mbox{E}[\lambda_{j,u}] 
	\mbox{Cov}[ \varepsilon_{i,t}, \varepsilon_{j,u}],  
\end{align*} 
for any $t,u$ and $i \neq j$ or $t \neq u$ and all $i,j$. Effectively the autocovariance function of the hidden process 
is identical to the autocovariance function of the standardized process $Y_{i;t}/ \mbox{E}[\lambda_{i,t}]$. 
\citet{WangandWang(2018)} assume further  that $ \mbox{E}[a_{i,t}]=\exp({c}_{t}^{T}{\alpha}_{i})$ using the
previous notation. To reduce the dimensionality of   hidden process $\varepsilon_t$, it is assumed to satisfy $\varepsilon_t= \Gamma f_{t}$ as in \citet{Jungetal(2011)} but with $s=\dim(f_{t})$ unknown. Correlation is taken 
into account by this construction since the dynamics of $f_t$ drive the time-evolution of $\varepsilon_t$.
Inference  proceeds in two steps: (a) pseudo-maximum-likelihood
estimation for regression coefficients and (b) identification of common factor(s) utilizing  
eigenanalysis on a positive definite matrix.

In a related article, \citet*{Zhangetalnew(2017)} discuss a model which is based on the  the multivariate lognormal
mixture Poisson distribution (see Sec. \ref{subseq:mixed Poisson}) and allows for serial correlations by 
assuming that  the Poisson mean vector is  a latent process driven by a nonlinear autoregressive model. 
The authors employ  Monte Carlo Expectation Maximization algorithm together with
particle filtering and smoothing methods to develop inference. Similarly,  \citet{AlWahsandGussein(2020)} motivated by an application concerning asthma related visits to emergency rooms, 
consider a hidden autoregressive process which drives the dynamics 
of  a positively correlated bivariate time series of counts whose conditional distribution is assumed 
to be  the multivariate Poisson distribution \eqref{eq:mult Poisson distribution}. The authors use a Bayesian 
data cloning approach to compute maximum likelihood estimators and their standard errors.

From a fully Bayesian point of view, \citet{Aktekinetal(2018)} (see also \citet{Gamermanetal(2013)}) assume that $Y_{i;t}$ are independent,
conditionally on univariate parameters $\alpha_{i}$ and  a process $\varepsilon_t$, Poisson distributed with mean 
$\alpha_{i} \theta_t$. The parameter $\alpha_{i}$ are individual specific rates  and $\varepsilon_t$ is a common 
process that drives the dynamics of the observed process and satisfies $\varepsilon_t=(\varepsilon_{t-1}/\gamma) d_{t}$, 
where $\gamma \in (0,1)$ and $d_{t}$ are independent Beta random variables with suitable parameters.  The authors 
study, in addition, a negative binomial model and they implement inference  by  particle learning
algorithm.  Another fully Bayesian approach is that of \citet{BerryandWest(2020)} who introduce models, within the 
framework of dynamic GLM (see \citet{WestandHarrison(1997)}), that allow use of time-varying covariates for 
binary and Poisson conditionally distributed time series. The recent review article by \citet{West(2020)} gives 
further insight for Bayesian modeling of multivariate count time series. Other works,  along these lines include 
\citet{Serhiyenko(2015)},  \citet{Ravishankeretal(2014)}, \citet{Ravishankeretal(2015)}. The previous articles 
and the recent work of \citet{Davisetal(2021)} give further references   and list other approaches.

\section{Observation-Driven Models}
\label{sec:Observation-Driven Models}

In this section, we discuss observation-driven models, that is processes whose  dynamics are driven by past observations plus noise.  A convenient  model would postulate a multivariate conditional count distribution to the observed process such
that likelihood inference is feasible. But the discussion in Sec. \ref{seq:distribitions} illustrated
the obstacles of choosing appropriate count
distribution.  In this section we will be studying  the   GLM approach. It will be argued that  
this framework  generalizes  the traditional ARMA methodology (see \citet{ShumwayandStoffer(2011)} for example)   to the count time series framework. 
Model fitting is based, in general,   on quasi-likelihood inference; \citet{Godambe(1991)} \citet{Heyde(1997)}; therefore testing, diagnostics and all type of likelihood arguments are directly applicable for this case. 

\subsection{Linear Models}

To initiate the discussion, consider the  standard  VAR(1) model, but in the context of a multivariate  Poisson autoregression
as it is  discussed next.  Denote by ${\cal F}_{t}$ the $\sigma$--field generated by all past values  
of the process  $\{ Y_{s}, ~s \leq t \}$. 
Let   $\{  \lambda_{t} = (\lambda_{i,t}), ~i=1,2,\ldots,d, t \in \mathbb{Z}  \}$ be the
corresponding $d$-dimensional intensity process, vis. $\mbox{E}[Y_t \mid {\cal F}_{t-1}]=\lambda_{t}$. 
The univariate linear autoregressive model  discussed by 
\citet{RydbergandShephard(2000)},  \citet{Heinen(2003)}, \citet{Ferlandetal(2006)} and \citet{Fokianosetal(2009)}, among others, 
serves as basic building block to construct a multivariate Poisson linear VAR(1) process  by defining 
\begin{equation}
	Y_{i, t} \mid {\cal F}_{t-1}  ~~\sim~~ \mbox{marginally Poisson}(\lambda_{i,t}), ~~~~~\lambda_{t}= \omega+B_{1}Y_{t-1},
	\label{eq:poisson inarch(1)}
\end{equation}
where ${\omega }$ is a $d$-dimensional vector and $B_{1}$ is a  $d \times d$ unknown  matrices. The elements of
$\omega$ and $B_{1}$ are assumed to be positive such that $\lambda_{i,t}> 0$. for all $i$ and $t$.
It is instructive to consider  \eqref{eq:poisson inarch(1)} in more detail. For the simple
case  $d=2$,  it implies that  
\begin{eqnarray*}
	\lambda_{1t} & = &  \omega_{1} + b_{1}^{11} Y_{1,(t-1)}+ b_{1}^{12} Y_{2,(t-1)}, \\
	\lambda_{2t} & = &  \omega_{2} + b_{1}^{21} Y_{1,(t-1)}+ b_{1}^{22} Y_{2,(t-1)},
\end{eqnarray*}
where $\omega_{i}$ is the $i$'th element of $\omega$ and $b_{1}^{ij}$ is the $(i,j)$th element of
$B_{1}$. Then setting $b_{1}^{12}=0$ implies that  the past values of $Y_{2,t}$ do
not affect the evolution of $Y_{1,t}$. Similarly,  $b_{1}^{21}=0$ shows that  
past values of $Y_{1,t}$ do not affect the evolution of $Y_{2,t}$. These  arguments
extend naturally  to the case $d > 2$. 

Generating data using \eqref{eq:poisson inarch(1)} is accomplished by imposing  \eqref{eq:mult Poisson distribution}, for instance; see \citet{Liu(2012)} and  \citet{PedeliandKarlis(2013)}   for some  examples. But the discussion in Sec. \ref{seq:distribitions} shows the
challenges of fitting model \eqref{eq:poisson inarch(1)} to data assuming  a bivariate Poisson (and more generally multivariate Poisson)
distribution. To overcome this challenge we appeal to copulas--but other suitable p.m.f are applicable--by introducing 
a joint distribution  constructed by utilizing   the data generating process described in Sec. 
\ref{subseq:Copulamodels} but  taking into account \eqref{eq:poisson inarch(1)}. The algorithm is repeated for completeness of presentation. Let  ${\lambda}_{0}$  be a  starting value and assume that $\omega$, $B_{1}$ are given. Then
\begin{enumerate}
	\item Let ${U}^{l}=(U_{1 ,l}, \ldots, U_{d,l})$ for $l=1,2,\ldots, K$, be a sample from a $d$-dimensional
	copula $C(u_{d}, \ldots, u_{d})$. Then $U_{i,l}$, $l=1,2,\ldots, K$ follow marginally the uniform distribution on $(0,1)$, for $i=1,2,\ldots,d$.
	\item Consider  the transformation
	%\begin{eqnarray*}
	$X_{i, l} = -{ \log U_{i, l}}/{\lambda_{i,0}}, ~~~i=1,2, \ldots, d.$
	%\end{eqnarray*}
	Then, the marginal distribution of $X_{i,l}$, $l=1,2,\ldots, K$
	is exponential with parameter $\lambda_{i,0}$, $i=1,2,\ldots,d$.
	\item
	%Define now (taking $K$ large enough)
	%\begin{eqnarray*}
	%$Y_{i,0} = \max_{1 \leq k \leq K} \left\{ \sum_{l=1}^{k} X_{i, l} \leq  1 \right\}, ~i=1,2,\ldots,d.$
	%\end{eqnarray*}
    {If $X_{i,1} >1$, set $Y_{i,0}=0$, otherwise 
		$Y_{i, 0} = \max \left\{K:~ \sum_{l=1}^{K} X_{i, l} \leq  1 \right\}, ~i=1,2,\ldots,d.$}
	Then ${Y}_{0}=(Y_{1,0}, \ldots, Y_{d,0})^T$ is marginally  a realization of a   Poisson process with parameter ${\lambda}_{0}$.
	\item Use \eqref{eq:poisson inarch(1)}  to obtain ${\lambda}_{1}$.
	\item Return back to step 1 to obtain ${ Y}_{1}$, and so on.
\end{enumerate}

\begin{figure}[h]
	\centering
	\includegraphics[width=0.8\linewidth, height=0.45\textheight]{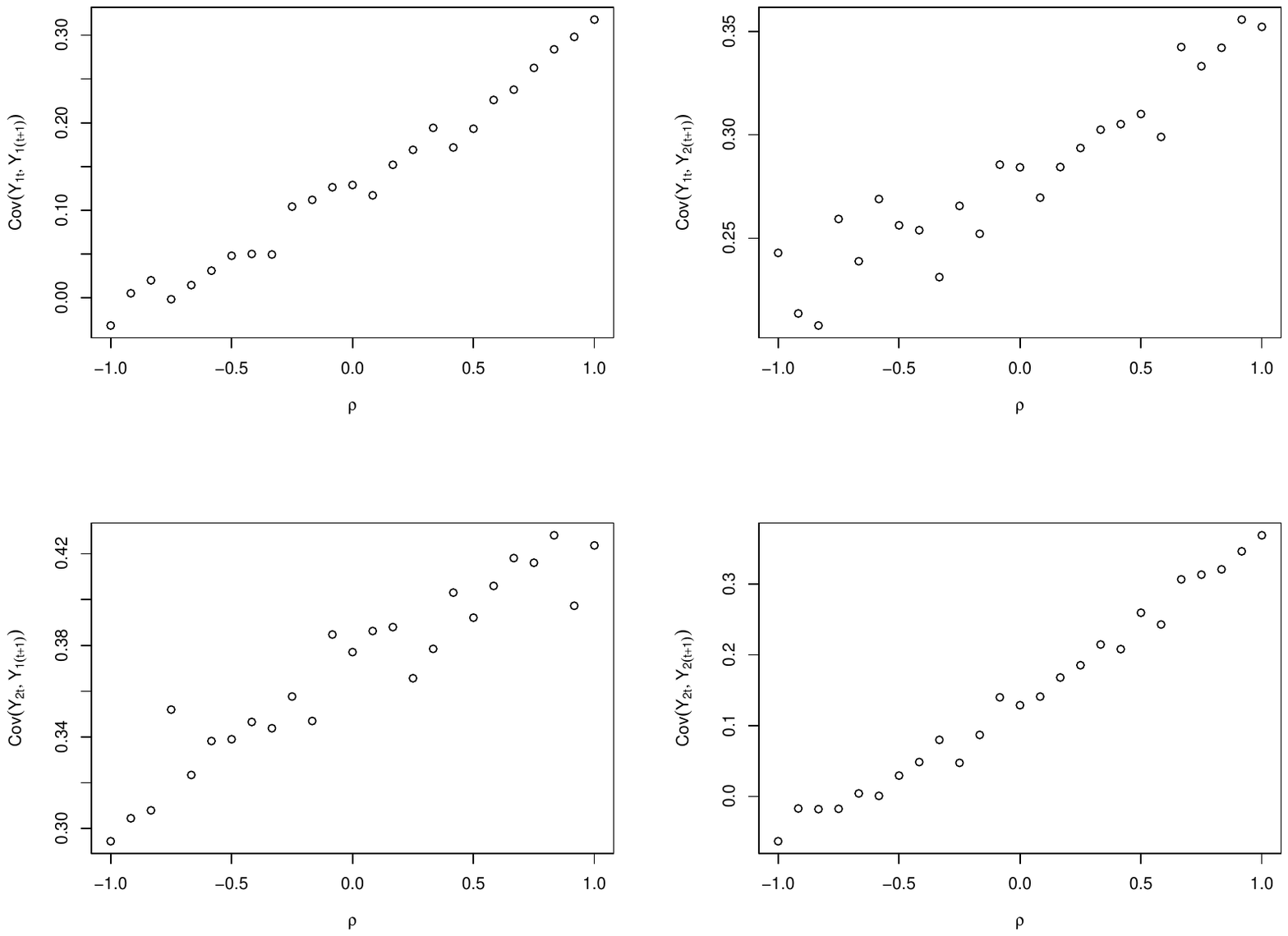}
	\caption{Lag 1 autocorrelation matrix of a bivariate count time series as a function of the Gaussian copula parameter $\rho$. Results are based on 5000 data points.}
	\label{fig:acfplot}
\end{figure}

Figure 2 shows plots of the sample autocorrelation matrix $\mbox{Cov}(Y_t, Y_{t+1})$ where $Y_{t}$ is a bivariate count time series
generated using the above algorithm and model \eqref{eq:poisson inarch(1)}  with  a Gaussian copula having parameter 
$\rho$.   The plot reinforces the point raised in Sec. \ref{subseq:Copulamodels} where it is noted that 
interpretation of the instantaneous correlation found in data  is related to the correlation of the vector
of waiting times and should be done with care. This simple example shows that, as $\rho$ varies between -1 and
1, all the lag 1 correlation functions do not exceed 0.45. But the plot also shows that this approach takes into 
account properly the correlation found in data {because by varying the parameter $\rho$ we obtain  different type of  autocorrelation matrices.}

It is worth pointing out that this approach  is different when compared to the work of \cite{HeinenandRegifo(2007)}. 
These authors replace the original counts by employing  the continued extension
method of \cite{DenuitandLambert(2005)}, as it was discussed in Sec. \ref{subseq:Copulamodels}. The continued extension method  of \cite{HeinenandRegifo(2007)} has been investigated in a simulation study by \cite{Nikoloulopoulos(2013)}.
Other copula-based models for multivariate count distributions with serial dependence are discussed in \citet[Ch. 8]{Joe(1997)}.

From now on, define \eqref{eq:poisson inarch(1)} as vector Integer Autoregressive Conditional Heteroscedastic  model of order 1, i.e. V-INARCH(1) model. The reason for choosing this terminology  will be explained below. 
Following identical arguments as those of \citet[Sec. 3.1]{Fokianos(2012)}, model \eqref{eq:poisson inarch(1)}
is rewritten  as 
\begin{eqnarray}
	Y_{t}  =   \lambda_{t} + (Y_{t}- \lambda_{t})
	=   \omega +B_{1}Y_{t-1} + \zeta_{t}, 
\label{eq:inarch1representation}
\end{eqnarray}
which shows that the values of  $Y_t$ depend on $Y_{t-1}$
plus the white noise sequence    $\{ \zeta_{t} \}$.  Indeed, if  $\{Y_t\}$    is assumed to be stationary, then 
it is easily shown that $\mbox{E}[\zeta_{t}]=0$, 	$\mbox{Var}[\zeta_{t}]=  \mbox{E}[\Sigma_t]$, where
$\Sigma_{t}=\mbox{Var}[Y_{t} \mid {\cal F}_{t-1}]$ and $ \mbox{Cov}(\zeta_{t}, \zeta_{t+k})=0$ for $k \in \mathbb{Z}$. 
The matrix $\Sigma_t$ is not  determined as in the univariate case whereby  
the conditional variance of $Y_t$ is $\lambda_t$ (with some abuse of notation). Model \eqref{eq:poisson inarch(1)} implies that the diagonal elements of $\Sigma_t$ are $\lambda_{i,t}$ but
the off-diagonal entries depend on the copula employed to generate data but  
are unknown because of the contemporaneous correlation between $Y_{i,t}$ and $Y_{j,t}$, $i \neq j$.
 
Because of the assumed stationarity, \eqref{eq:poisson inarch(1)}
shows that $\mbox{E}[Y_{t}]= \omega + B_{1} \mbox{E}[Y_{t-1}]$. Then 
$\mbox{E}[Y_{t}] = (I_d-B_{1})^{-1} \omega $,  provided that $\rho(B_1) <1$, where $\rho(.)$
denotes the spectral radius of a matrix.
%In other words $\det(I_d-z B_{1})  \neq 0$ for all $z$ 
%such that $|z| \leq 1$.  
Furthermore,  (see \citet[Ch.2]{Lutkepohl(2005)}) 
\begin{eqnarray}
	Y_{t} & = &  \omega +B_{1}Y_{t-1} + \zeta_{t} \nonumber  \\
	& = &   \omega +B_{1}( \omega+B_{1}Y_{t-2}+\zeta_{t-1})+\zeta_{t} \nonumber  \\
	& = &   (I_d+B_{1})\omega  + B_{1}^{2}Y_{t-2} + B_{1} \zeta_{t-1} + \zeta_{t} \nonumber  \\
	& = &  \cdots    \cdots    \cdots \cdots    \cdots    \cdots   \nonumber \\
	& = &    (1+B_{1}+B_{1}^{2}+\cdots B_{1}^{t})\omega + \sum_{i=0}^{t} B_{1}^{i} \zeta_{t-i}. \label{eq:onestepreplinear}
\end{eqnarray}
Therefore, as in the case of ordinary  VAR(1) model,  assuming that  $\rho(B_1) <1$, we obtain (in mean square sense) by \eqref{eq:onestepreplinear} and  for large $t$, the useful one-sided infinite order moving average representation
\begin{eqnarray}
	Y_{t} = (I_d-B_{1})^{-1} \omega  + \sum_{i=0}^{\infty} B_{1}^{i} \zeta_{t-i}.
	\label{aymptotic representationforlinear}
\end{eqnarray}
In addition
\begin{align*}
\Gamma_{Y}(h) \equiv \mbox{Cov}(Y_{t}, Y_{t+h}) = \sum_{i=0}^{\infty}
 B_{1}^{i+h} \mbox{E}[\Sigma_t] (B^{i})^{T}, ~~~~ h \geq 0,
\end{align*}
Several other  results  are readily available because of  \eqref{eq:inarch1representation}, see
\citet{Lutkepohl(2005)} or \citet{Tsay(2014)}.

Generalizations of \eqref{eq:poisson inarch(1)}, such as the V-INARCH($p$) model
\begin{align*}
	\lambda_{t}=\omega+\sum_{i=1}^{p} B_{i} Y_{t-i},
\end{align*}
or the vector V-INGARCH($p$, $q$) (where "G" stands for Generalized)
\begin{align}
	Y_{i,t} \mid {\cal F}_{t-1}   ~~ \mbox{is marginally} ~~ \mbox{Poisson}(\lambda_{i, t}),  ~~
	{ \lambda}_{t}  =   \omega  + 	\sum_{i=1}^{p} B_{i} Y_{t-i}+  \sum_{j=1}^{q} A_{j} \lambda_{t-j},
	\label{eq:linear model mult}
\end{align}
where   $ ({A_{j}})_{j=1}^{q}$, $({B}_{i})_{i=1}^{p}$ are $d \times d$ unknown  matrices
and all the elements of $\omega$,  $ ({A_{j}})_{j=1}^{q}$, $({B}_{i})_{i=1}^{p}$ are  positive such that $\lambda_{i,t}> 0$. for all $i$ and $t$, are studied along the previous arguments. 
The abbreviation "INGARCH" for model   \eqref{eq:linear model mult} just indicates 
its  structural connection  to ordinary   GARCH model, \citet{Bollerslev(1986)},
because  each component of the vector process
$Y_{t}$ is distributed as a Poisson random variable. But
the mean of a Poisson random variable equals its variance; therefore the  structure of
\eqref{eq:linear model mult}  bears some resemblance to that of  
a multivariate GARCH model, see \citet{FrancqandZakoian(2010)}. Though the term "V-INGARCH"
does not reflect accurately the true data generating process it will be used as a generalization 
of the terminology introduced  for univariate models by \citet{Ferlandetal(2006)}.

It  is proved that   
\eqref{eq:linear model mult} is  a VARMA($\max(p,q)$, $q$) process. Recall   that $\zeta_{t}={Y}_{t}-{\lambda}_{t}$.
Then assuming first order stationarity of $Y_{t}$ and taking expectations on both sides of \eqref{eq:linear model mult}, we obtain
that $\mbox{E}[Y_t]=(I_d-\sum_{i=1}^{p} A_{i}- \sum_{j=1}^{q} B_j)^{-1} \omega $, provided that $\rho(
\sum_{i=1}^{p} A_{i} +\sum_{j=1}^{q} B_j) <1$. Then, by manipulating \eqref{eq:inarch1representation} but for model \eqref{eq:linear model mult}, it is easily shown that 
\begin{align}
	\biggl({Y}_{t}- \mbox{E}[Y_t] \biggr)=  
	\sum_{i=1}^{\max(p,q)} \biggl({A}_{i}+{B}_{i} \biggl) \biggr( {Y}_{t-i}- \mbox{E}[Y_t] \biggr)+ \zeta_t - \sum_{j=1}^{q} {A}_j \zeta_{t-j},
	\label{eq:VARMA for linear model}
\end{align}
where we set $A_{i}=0_d$ if $q <p$ for $i=q+1, \ldots,p$ or $B_{i}=0_d$ if $q>p$ for $i=p+1, \ldots, q$. 
In the case of $p=q=1$ 
the  one-sided MA($\infty$) is given by ${Y}_{t}= \mathbold{\mu}+\sum_{j=0}^{\infty} \Phi_{j} \zeta_{t-j}$
with ${\Phi}_{0}= {I}_{d}$ and ${\Phi}_{j}= (A_{1}+B_{1})^{j-1} B_{1}$, for $j \geq 1$, a fact that shows for any $h >0$
\begin{align*}
	\Gamma_{Y}(h)=
	\sum_{j=0}^{\infty}
	\bigl( {A}_1+ {B}_1 \bigr)^{j+h-1} {B}_1 \mbox{E}[{\Sigma}_{t}] {B}_{1}^{T} \bigl( {A}_{1}^{T}+ {B}_{1}^{T} \bigr)^{j+h-1}, 
\end{align*}
by using properties of the linear multivariate processes.

\subsection{Log-linear Models}

The log-linear model we consider is  the  multivariate analogue of the univariate  log-linear model proposed by \citet{FokianosandTjostheim(2011)}. In a more general form,  assume that for each  $i=1,2,\ldots,d$
\begin{eqnarray}
	Y_{i, t} \mid {\cal F}_{t-1} ~~ \mbox{is marginally} ~~ \mbox{Poisson}(\lambda_{i,t}),~~
	{\nu}_{t}  =   \omega  + 	\sum_{i=1}^{p} B_{i} \log(Y_{t-i} +1_d)+  \sum_{j=1}^{q} A_{j} \nu_{t-j}
	\label{eq:log-linear model mult}
\end{eqnarray}
where ${ \nu}_{t} \equiv \log  \lambda_{t}$
is defined component wise (i.e. $\nu_{i,t}= \log \lambda_{i,t}$) and ${ 1}_{d}$ denotes the $d$--dimensional
vector which consists of ones. For this model, there is no need to  impose any constraints on the matrix  coefficients. 
Additionally, the log-linear model accommodates covariates much easier than the linear model which requires that any such inclusion has to satisfy $\lambda_{i,t} >0$. In the case of \eqref{eq:log-linear model mult} though,   if ${Z}_{t}$ is a covariate 
vector of dimension $d$, then the second equation of \eqref{eq:log-linear model mult}
becomes  ${ \nu}_{t}  =  \omega  + 	\sum_{i=1}^{p} B_{i} \log(Y_{t-i} +1_d)+  \sum_{j=1}^{q} A_{j} \nu_{t-j} +C Z_{t}$ for a $d\times d$ matrix ${C}$.  Interpretation of model parameters for the log-linear model 
\eqref{eq:log-linear model mult} is identical to the case of the linear model but in terms of the vector
process $\nu_{t}$.

It is more challenging to derive formulas for the mean and autocovariances of model \eqref{eq:log-linear model mult}.
However, some approximations are possible by considering the process 
${W}_{t} \equiv \log \bigl( {Y}_{t} + {1}_{d} \bigr)$. Indeed, define now ${\zeta_t}= {W}_{t}-{\nu}_{t}$
and use the results of \cite{FokianosandTjostheim(2011), Fokianosetal(2019)} to see that
${W}_{t}$ is \emph{approximated} by a VARMA  model of the form 
\begin{align*}
	\biggl({W}_{t}- \mbox{E}[W_t] \biggr)=  
	\sum_{i=1}^{\max(p,q)} \biggl({A}_{i}+{B}_{i} \biggl) \biggr( {W}_{t-i}- \mbox{E}[W_t] \biggr)+ \zeta_t - \sum_{j=1}^{q} {A}_j \zeta_{t-j},
\end{align*}
similar to \eqref{eq:VARMA for linear model}. An approximate formula for the sequence
of autocovariance matrices for  ${W}_{t}$  (but not for $Y_t$) is then derived  but with suitable adjustments.
{This representation should be used cautiously because it is approximate and it can be applied for
developing a model for 	$W_t$ (using standard time series methodology)  but not for $Y_t$.  }

\begin{Remark} \rm
Stability conditions for model  \eqref{eq:linear model mult}  have been developed
by \citet{Liu(2012)} (under the framework of  multivariate Poisson distribution \eqref{eq:mult Poisson distribution}) and 
\citet{Fokianosetal(2019)} under the copula construction as outlined in Sec. \ref{subseq:Copulamodels} for the case $p=q=1$.
Recently \citet{Truquet(2019)} have improved these conditions again considering the copula-based data generating 
process, as outlined before.  Without introducing any further notation, we note that the condition $\rho(\sum_{i} A_{i}+\sum_{j} B_{j}) < 1$
guarantees stability of the process. For the log-linear model \eqref{eq:log-linear model mult} the desired conditions
are more complicated; see \citet{Fokianosetal(2019)} who consider 
the case $p=q=1$ and prove that either 
$\|| A_{1}\||_{2} + \||B_{1} \||_{2} < 1$ or  $\|| {A}_1 \||_{1} + \|| {B}_{1} \||_{1} < 1$, where  $\|| {A} \||_{d}= \max_{\|{ x}\|_{d}=1} \| A {x} \|_{d}$, guarantee ergodicity of the process. 
Related stability conditions are discussed in \citet{Truquet(2019)}.

The main notions used  to derive such conditions are those of Markov chain theory (\citet{MeynandTweedie(1993)}), weak dependence
(\citet{DoukhanandLouhichi(1999)}, \citet{Dedeckeretal(2007)})  and  convergence of  backward iterations of random maps
(\citet{WuandShao(2005)}). Following the discussions of \citet{Neumann(2010)} and \citet{Tjostheim(2012), Tjostheim(2015)}, the main difficulty
is that the process itself consists of integer valued random variables; however the mean process takes values on the positive real line and
therefore it is quite challenging to prove   stability of the joint  process (see also \citet{Andrews(1984)}).
The study of theoretical properties of univariate  models was initiated by the perturbation method suggested in \citet{Fokianosetal(2009)} and
was further developed in \cite{Neumann(2010)} (using the notion of $\beta$-mixing), \citet{Doukhanetal(2011)} (weak dependence approach),  \citet{Woodardetall(2010)} and
\citet{Doucetal.(2012)} (Markov chain theory without irreducibility assumptions) and \citet{Liu(2012)}, \citet{Wangetal(2014)} (based on the theory of $e$-chains).
\end{Remark}

\begin{Remark} { \rm
Models  \eqref{eq:linear model mult} and \eqref{eq:log-linear model mult}	are  related to Hawkes processes ( \cite{Hawkes(1971a), Hawkes(1971b)}) because they  can be obtained by suitable  discretization of the continuous  time process. This connection has been  explored in detail by  \citet{Kirchner(2016)} for univariate models in the context of INAR($\infty$)  process which in turn is related to the linear model (see \citet{Ferlandetal(2006)}). Hawkes processes have been found useful in modeling and inference in several scientific areas--a review is out of the scope of this paper. In particular, multivariate Hawkes processes have been employed extensively in finance; see \citet{Embrechtsetal(2011)}
and \citet{Bacryetal(2015)} among others.
}
	
\end{Remark}

\subsection{Quasi-Likelihood Inference}

Suppose that $\{ Y_{t}, t=1,2,\ldots,n \}$  is an available sample  from a  count time series and 
for the sake of presentation assume model  \eqref{eq:linear model mult} for $p=q=1$. 
Inference  is analogously developed  to the case of  log--linear model and for $p,q >1$.
Denote  
by ${\theta }=({d}^{T}, \vect^{T}(A_1), \vect^{T}(B_1))$,  $\dim(\theta) \equiv \kappa= d(1+2d)$.
Following \citet{Fokianosetal(2019)}, the estimation problem is approached  by employing  the theory of estimating functions. 
Consider  the following conditional quasi--likelihood function, given a starting value  $\lambda_{0}$,   for the parameter vector  ${\theta }$,
\begin{eqnarray*}
	L({\theta}) = \prod_{t=1}^{n} \prod_{i=1}^{d}\Bigl\{ \frac{ \exp(-\lambda_{i,t}({ \theta}))\lambda_{i,t}^{y_{i,t}}({ \theta})}{y_{i,t}!} \Bigr\},
\end{eqnarray*}
which is identical  to  consider \eqref{eq:linear model mult} assuming 
independence among time series.  This is strong  assumption yet it  simplifies  computation of estimators and their respective standard errors. At the same time, it guarantees consistency and asymptotic normality of the
maximizer. Furthermore, the dependence structure in \eqref{eq:linear model mult}
and \eqref{eq:log-linear model mult} is taken into account  through because of  
the dependence of the likelihood function  on the matrices ${A_1}$ and ${B_1}$; 
see \citet{Fokianosetal(2019)} for more.  
The quasi  log-likelihood function is equal to
\begin{eqnarray*}
	l({ \theta}) = \sum_{t=1}^{n} \sum_{i=1}^{d} \Bigl( y_{i,t} \log \lambda_{i,t}({ \theta})-\lambda_{i,t}({ \theta}) \Bigr).
	\label{quasi-loglikelihood}
\end{eqnarray*}
We denote by  $\widehat{{\theta}}\equiv \arg \max_{{\theta}} l({ \theta}), $ the QMLE of ${\theta}$.
The score function is given by
\begin{eqnarray}
	S_{n}( \theta)= \sum_{t=1}^{n} \sum_{i=1}^{d} \Bigl( \frac{y_{i,t}}{\lambda_{i,t}({\theta})}-1 \Bigr)
	\frac{\partial \lambda_{i,t}( \theta)}{\partial {\theta}}
	=  \displaystyle \sum_{t=1}^{n} \frac{ \partial {{\lambda}}^{T}_{t}({\theta})}{ \partial {{\theta}}} { D}^{-1}_{t}
	({ \theta})\Bigl({ Y}_{t}- {{\lambda}}_{t}({{\theta}})\Bigr)
	\label{score independnence}
\end{eqnarray}
where ${\partial {{\lambda}}_{t}}/{ \partial {{\theta}^{T}}}$ is a $d \times \kappa$ matrix and ${D}_{t}$ is the $d \times d$
diagonal matrix with the $i$'th diagonal element equal to $\lambda_{i,t}({\bf \theta})$, $i=1,2,\ldots,p$. Furthermore
\begin{eqnarray*}
	\frac{ \partial {{\lambda}}_{t}}{ \partial {{d}^{T}}}  & = & { I}_{d} + { A_1} \frac{ \partial {{\lambda}}_{t-1}}{ \partial {{d}}^{T}}, \nonumber\\
	\frac{ \partial {{\lambda}}_{t}}{ \partial \vect^{T}({ A_1})}    & = & ({ \lambda}_{t-1} \otimes { I}_{d})^{T} + { A_1}
	\frac{ \partial {{\lambda}}_{t-1}}{ \partial \vect^{T}({ A_1}) },  \label{recursions linear} \\
	\frac{ \partial {{\lambda}}_{t}}{ \partial \vect^{T}({ B_1})}    & = & ({ Y}_{t-1} \otimes { I}_{d})^{T} + { A_1}
	\frac{ \partial {{\lambda}}_{t-1}}{ \partial \vect^{T}({ B_1}) }. \nonumber
\end{eqnarray*}
The   Hessian matrix is given by
\begin{eqnarray}
	{ H}_{n}({\theta}) & = &
	\sum_{t=1}^{n} \sum_{i=1}^{p} \frac{y_{i;t}}{\lambda_{i, t}^{2}({\theta})}
	\frac{\partial \lambda_{i,t}({\theta})}{\partial {\theta}}
	\frac{\partial \lambda_{i,t}({\theta})}{\partial {\theta}^{T}}-
	\sum_{t=1}^{n} \sum_{i=1}^{p} \Bigl( \frac{y_{i,t}}{\lambda_{i,t}({\theta})}-1 \Bigr)
	\frac{\partial^{2} \lambda_{i,t}( \theta)}{\partial {\theta} \partial {\theta}^{T}}.
	\label{Hessian}
\end{eqnarray}
Therefore,  the conditional information matrix is equal to
\begin{eqnarray}
	{ G}_{n}({\theta})& = &
	\sum_{t=1}^{n}   \frac{ \partial {{\lambda}}^{T}_{t}({\theta})}{ \partial {{\theta}}} { D}^{-1}_{t}({\theta})
	{\Sigma}_{t}({\theta})
	{ D}^{-1}_{t}({\theta}) \frac{ \partial {{\lambda}}_{t}({\theta})}{ \partial {{\theta}^{T}}},
	\label{information}
\end{eqnarray}
where the matrix ${\Sigma}_{t}(\cdot)$ denotes the \emph{true} covariance matrix of the vector ${Y}_{t}$. In case
that the process $\{{Y}_{t}\}$ consists of uncorrelated components then $ {\Sigma}_{t}({\theta})=
{D}_{t}({\theta})$. In the case that of $A_{1}$ being a diagonal matrix then $\hat{\theta}$ is computed by equation by equation using existing software.

Under suitable conditions, \citet{Fokianosetal(2019)} show that 
\begin{equation*}
	\sqrt{n}(\widehat{{\theta}}-{\theta}_0) \xrightarrow{\text{d}} N(0, { H}^{-1} { G} { H}^{-1})
\end{equation*}
where the matrices ${H}$ and ${G}$ are defined by  the limits (in probability) of \eqref{Hessian} and \eqref{information}, respectively. The same result is true for the log-linear model \eqref{eq:log-linear model mult}; details are 
omitted.  To estimate the copula parameter, it is desirable to compare the  conditional
distribution of ${Y}_{t} \mid {\lambda}_{t}$ to that  of ${Y}^{*}_{t} \mid {\lambda}_{t}$, where
${Y}^{*}_{t}$ is a count time series 
generated by a suitable choice of a copula. There are several ways of comparing such distributions and this topic
is still under investigation. In \citet{Fokianosetal(2019)}  an initial approach,  based on the newly developed concept of local Gaussian correlation (see \citet{Berentsenetal(2014)})  was  shown to be satisfactory. But the problem of estimating the copula parameter remains unexplored; see \citet{TruquetandDebaly(2021)}
for recent progress in the framework of mixed time series models.

\begin{Remark} \rm 

Equation \eqref{score independnence} motivates  a more general framework  that can be applied  to 
the analysis of multivariate count time series modes. A natural generalization, is to consider
the following estimating functions 
\begin{eqnarray}
	S_{{ v}}({ \theta}) = \sum_{t=1}^{n}  \frac{ \partial {{\lambda}}^{T}_{t}({\theta})}{ \partial {{\theta}}}
	{ V}^{-1}_{t}(\rho, { \lambda}_{t}({ \theta}))
	\Bigl({ Y}_{t}- {{\lambda}}_{t}( \theta)\Bigr),
	\label{score}
\end{eqnarray}
where the notation is completely analogous to \eqref{score independnence} and ${ V}_{t}(\rho, { \lambda}_{t}({ \theta}))$
is a  $d \times d$ "working" conditional
covariance matrix which depend upon the process $\{ {\lambda}_{t} \}$ and possibly some other parameters $\rho$.
Several choices for the working conditional covariance matrix are available in the literature; we list some possibilities.
If  ${V}={ I}_{d}$, then \eqref{score} corresponds to a  least squares minimization problem for
estimating ${ \theta}$. If $V=D_{t}$ then we obtain  \eqref{score independnence}. More generally, the choice 
	$$
	{ V}(\rho, {{\lambda}})=\left(       \begin{array}{cccc}
		\lambda_{1, t} & \rho_{12} \sqrt{\lambda_{1,t}}\sqrt{\lambda_{2,t} } & \cdots  & \rho_{1d} \sqrt{\lambda_{1,t}}\sqrt{\lambda_{d,t}}  \\
		\rho_{12} \sqrt{\lambda_{1,t}}\sqrt{\lambda_{2,t}}  &  \lambda_{2,t} &  \cdots  & \rho_{2d} \sqrt{\lambda_{2,t}}\sqrt{\lambda_{d,t}}   \\
		\cdots  &  \cdots  &  \cdots  & \cdots  \\
		\rho_{1p} \sqrt{\lambda_{1,t}}\sqrt{\lambda_{p,t}}   & \rho_{2p} \sqrt{\lambda_{2,t}}\sqrt{\lambda{p,t}} & \cdots & \lambda_{d,t} \\
	\end{array}
	\right)
	$$
yields to a constant conditional correlation type of model for multivariate count time series, see \citet{Terasvirtaetal(2010)}, among others. This topic deserved  further research; a possible method might rely on  
the   \citet{FrancqandZakoian(2016)} who consider estimation of  multivariate volatility models equation by equation.
\end{Remark}

\subsection{High-Dimensional  Models}
\label{highdimModels}

{The advent of technology to economics, biological and social sciences, 
has given rise to interesting and exciting application of 
high-dimensional time series models. Some examples include multiple transactions of several stocks, 
gene regulatory network reconstruction from  time course gene expression data, brain connectivity
analysis and others. Such applications have revived methodology which is useful for the purpose of
modeling and inference. Though the concept of sparsity attracted a lot of attention over the last two decades and
proved its usefulness for modeling and inference (see \citet{Hastieetal(2015)} among others) research on time series 
methods is still in progress. Because such a review is out of the scope of this article, we refer the reader to recent work
by \citet{BasuandMatteson(2021)} who provide an overview about several methods, in the context of large autoregressions and stochastic regression, and  \citet{Hallinetal(2020)}  who provide a concise overview of  factor models.
\newline
In the rest, we outline a recent methodological contribution related to  inference for high-dimensional count time series observation driven models; see \citet{Halletal(2019}. Those authors assume the pure autoregressive model 
\begin{eqnarray}
	Y_{i, t} \mid Y_{t-1}  ~~ \mbox{Poisson}(\lambda_{i,t}),~~
	{\nu}_{t}  =   \omega  + 	B_{1} Y_{t-1} 
	\label{eq:log-linear high-dim}
\end{eqnarray}
where ${ \nu}_{t}$ defined as in eq. \eqref{eq:log-linear model mult}. The constant term $\omega$ is assumed to be known 
and the $d\times d$ matrix $B_{1}$  belongs to a compact subset, say ${\cal B}$ of the set of all $d \times d$ matrices with real elements such that $\Vert B_{1} \Vert_{0} \equiv \sum_{l=1}^{d} \sum_{m=1}^{d} 1( \vert B_{1}^{(l,m)} \vert) \leq s$. The notation $1(.)$ denotes the indicator function. 
 Comparing \eqref{eq:log-linear high-dim} to \eqref{eq:log-linear model mult}  we note that the restriction 
of $B_{1}$  belonging  to a compact set assures the stability of the joint process $Y_{t}$ because the components
of $Y_{t-1}$ are  unbounded, in general. So when some regression coefficients are positive  the conditional expectation
of the response given the past of the process tends to grow in an exponential rate.
To estimate $B_{1}$ when $d$ is much larger than $n$, the authors propose the $l_{1}$ regularized QMLE  defined by 
\begin{align*}
\hat{B}_{1} = \underset{B_{1} \in {\cal B}}{ \arg \max} \frac{1}{n}	 \sum_{t=1}^{n} \sum_{i=1}^{d} 
\Bigl( y_{i,t} \nu_{i,t}( B_{1}) - \exp(\nu_{i,t}(B_{1})) \Bigr) + \tilde{\lambda} \sum_{l=1}^{d} \sum_{m=1}^{d} 
\vert B_{1}^{(l,m)} \vert , 
\end{align*}
where $\tilde{\lambda}$ is a regularization parameter and $\nu_{i,t}(.)$ is defined by \eqref{eq:log-linear high-dim}. The authors 
study mean square error bounds for the proposed estimators and show that they are closely connected with the bounds obtained in the n
Gaussian case. Further work along these lines was developed (for AR($p$) type models) by \citet{Panditetal(2020)} where the interested reader can obtain more references. 
\newline 
In another related work, \citet{armillotta_2021}  studied  network autoregressive models for high-dimensional count 
time series with a fixed neighborhood structure.  Assessing  the effect  of a network to  multivariate time series processes 
has attracted considerable attention over the last years. In particular, \citet{zhu2017} proposed  a Network Autoregressive model (NAR) and studied least squares inference under two asymptotic regimes (a) with increasing time sample size $n$ and fixed network dimension $d$ and (b) with both $n,d$ increasing.  These ideas are extended to high-dimensional count time series by 
\citet{armillotta_2021} who propose linear  and log-linear Poisson network autoregressions (PNAR) for count processes and by establishing the two related types of asymptotic inference for the QMLE as discussed before.
}

\section*{Acknowledgments}
Many thanks to three anonymous reviewers and  M. Armillotta who provided several comments that improved the
original submission.  This work has been funded by the European Regional development Fund and the Republic of Cyprus through the Research and innovation Foundation, under the project INFRASTRUCTURES/1216/0017 (IRIDA).

%{\small \bibliographystyle{chicago}
%\bibliography{C:/Work/Research/TimeSeries/bookbib}}

\end{document}